\documentclass[italian,english]{article}
\usepackage[latin9]{inputenc}
\usepackage{color}
\usepackage{amsthm}
\usepackage{amsmath}
\usepackage{amssymb}

\theoremstyle{plain}
\newcommand{\lyxaddress}[1]{
\par {\raggedright #1
\vspace{1.4em}
\noindent\par}
}

\usepackage{babel}

\begin{document}

\title{\textbf{Information on the inflaton field from the spectrum of relic
gravitational waves }}

\author{\textbf{Christian Corda }}

\maketitle

\lyxaddress{\begin{center}
Associazione Scientifica Galileo Galilei, Via Pier Cironi 16 - 59100
PRATO, Italy 
\par\end{center}}

\begin{center}
\textit{E-mail address:} \textcolor{blue}{cordac.galilei@gmail.com}
\par\end{center}
\begin{abstract}
After a review of a traditional analysis, it is shown a variation
of a more recent treatment on the spectrum of relic gravitational
waves (GWs). Then, a connection between the two different treatments
will be analysed. Such a connection permits to obtain an interesting
equation for the inflaton field. This equation gives a value that
agrees with the slow roll condition on inflation.
\end{abstract}

\section{Introduction}

The scientific community aims in a first direct detection of GWs in
next years (for the current status of GWs interferometers see \cite{key-1})
confirming the indirect, Nobel Prize Winner, proof of Hulse and Taylor
\cite{key-2}. 

Detectors for GWs will be important for a better knowledge of the
Universe and either to confirm or to rule out, in an ultimate way,
the physical consistency of General Relativity, eventually becoming
an observable endorsing of Extended Theories of Gravity, see \cite{key-3}
for details.

It is well known that an important potential source of gravitational
radiation is the relic stochastic background of GWs, see \cite{key-4}
for a recent review. The potential existence of such a relic stochastic
background arises from general assumptions. In fact, it derives from
a mixing between basic principles of classical theories of gravity
and of quantum field theory. The zero-point quantum oscillations,
which produce relic GWs, are generated by strong variations of the
gravitational field in the early universe. Then, the detection of
relic GWs is the only way to learn about the evolution of the very
early universe, up to the bounds of the Planck epoch and the initial
singularity \cite{key-4,key-5,key-6,key-7}. The importance of this
\textit{gravity's rainbow} in cosmological scenarios has been discussed
in an elegant way in \cite{key-8}. 

The model derives from the inflationary scenario for the early universe
\cite{key-6,key-9}, which is tuned in a good way with the WMAP data
on the Cosmic Background Radiation (CBR) (in particular exponential
inflation and spectral index $\approx1$ \cite{key-10,key-11}). Recently,
the analysis has been adapted to extended theories of gravity too
\cite{key-4,key-12,key-13}.

After a review of a traditional analysis on the spectrum of relic
GWs following \cite{key-7,key-12,key-13}, in this paper a variation
a of more recent analysis that uses a conformal treatment is performed.
A connection between the two different treatments will be also analysed.
This connection permits to obtain an interesting equation for the
inflaton field. This equation gives a value that agrees with the slow
roll condition on inflation \cite{key-9,key-20}.

For a sake of simplicity, in this paper natural units are used, i.e.
$8\pi G=1$, $c=1$ and $\hbar=1$. The Latin indices run from 1 to
3, the Greek ones from 0 to 3.

Considering a stochastic background of GWs, it can be characterized
by a dimensionless spectrum \cite{key-7,key-12,key-13}\begin{equation}
\Omega_{gw}(f)\equiv\frac{1}{\rho_{c}}\frac{d\rho_{gw}}{d\ln f},\label{eq: spettro}\end{equation}

where \begin{equation}
\rho_{c}\equiv\frac{3}{8}H_{0}^{2}\label{eq: densita critica}\end{equation}

is the (actual) critical density energy, $\rho_{c}$ of the Universe,
$H_{0}$ the actual value of the Hubble expansion rate and $d\rho_{gw}$
the energy density of relic GWs in the frequency range $f$ to $f+df$.

The more recent values for the spectrum can be found in \cite{key-12,key-13,key-14,key-15}.

\section{A review of the traditional analysis on the relic GWs spectrum}

Following the traditional analysis in \cite{key-7,key-12,key-13}
it will be assumed that the universe is described by a simple cosmology
in two stages, an inflationary De Sitter phase and a radiation dominated
phase. Then, the line element of the spacetime is given by \cite{key-7,key-12,key-13} 

\begin{equation}
ds^{2}=a^{2}(\eta)[-d\eta^{2}+d\underline{x}+h_{\mu\nu}(\eta,\underline{x})dx^{\mu}dx^{\nu}],\label{eq: metrica}\end{equation}

and the calculation will be performed for the exponential inflationary
model that agrees with the WMAP data \cite{key-10,key-11}.

In the De Sitter phase ($\eta<\eta_{1}$) the equation of state is
$P=-\rho=const$, the scale factor is $a(\eta)=\eta_{1}^{2}\eta_{0}^{-1}(2\eta_{1}-\eta)^{-1}$
and the Hubble constant is given by $H_{ds}=\eta_{0}/\eta_{1}^{2}$
\cite{key-7,key-12,key-13}.

In the radiation dominated phase $(\eta>\eta_{1})$ the equation of
state is $P=\rho/3$, the scale factor is $a(\eta)=\eta/\eta_{0}$
and the Hubble constant is given by $H=\eta_{0}/\eta^{2}$ \cite{key-7,key-12,key-13}. 

Expressing the scale factor in terms of comoving time defined by \cite{key-16}

\begin{equation}
dt=a(t)d\eta\label{eq: tempo conforme}\end{equation}
one gets

\begin{equation}
a(t)\varpropto\exp(H_{ds}t)\label{eq: inflazione}\end{equation}
during the De Sitter phase, and

\begin{equation}
a(t)\varpropto\sqrt{t}\label{eq: dominio radiazione}\end{equation}
 during the radiation dominated phase. In order to obtain a solution
for the horizon and flatness problems one needs \cite{key-9}

\begin{center}
\begin{equation}
\frac{a(\eta_{0})}{a(\eta_{1})}>10^{26}.\label{eq: inflazionata}\end{equation}

\par\end{center}

This value is the equivalent of 60 e-foldings, where one e-folding
is defined as the amount of time for $a$ to grow by a factor of $e$
\cite{key-9,key-20}.

The relic GWs are the weak perturbations $h_{\mu\nu}(\eta,\underline{x})$
of the metric (\ref{eq: metrica}). By considering, for example, the
$+$ plus polarization of GWs \cite{key-16}, in terms of the conformal
time $\eta$ it is \cite{key-7,key-12,key-13} 

\begin{equation}
h_{+}(\eta,\underline{k},\underline{x})=X(\eta)\exp(\underline{k}\times\underline{x}),\label{eq: phi}\end{equation}

where $\underline{k}$ is a constant wave vector.

By putting $Y(\eta)=a(\eta)X(\eta)$  and with the standard linearized
calculation in which the connections (i.e. the Cristoffel coefficients),
the Riemann tensor, the Ricci tensor and the Ricci scalar curvature
are found, from Friedman linearized equations one gets that the function
$Y(\eta)$ satisfies the equation \cite{key-7,key-12,key-13}

\begin{equation}
\frac{d^{2}Y}{d\eta^{2}}+(k^{2}-\frac{1}{a}\frac{d^{2}a}{d\eta^{2}})Y=0\label{eq: Klein-Gordon}\end{equation}
Clearly, this is the equation for a parametrically perturbed oscillator.

The solutions of eq. (\ref{eq: Klein-Gordon}) give the solutions
for the function $X(\eta)$, that can be expressed in terms of elementary
functions simple cases of half integer Bessel or Hankel functions
\cite{key-7,key-12,key-13} in both of the inflationary and radiation
dominated eras:

for $\eta<\eta_{1}$ \begin{equation}
X(\eta)=\frac{a(\eta_{1})}{a(\eta)}[1+H_{ds}\omega^{-1}]\exp-ik(\eta-\eta_{1}),\label{eq: ampiezza inflaz.}\end{equation}

for $\eta>\eta_{1}$ \begin{equation}
X(\eta)=\frac{a(\eta_{1})}{a(\eta)}[\alpha\exp-ik(\eta-\eta_{1})+\beta\exp ik(\eta-\eta_{1}),\label{eq: ampiezza rad.}\end{equation}
where $\omega=k/a$ is the angular frequency of the wave (that is
function of the time because of the constancy of $k=|\underline{k}|$),
$\alpha$ and $\beta$ are time-independent constants which can be
obtained by demanding that both $X$ and $dX/d\eta$ are continuous
at the boundary $\eta=\eta_{1}$ between the inflationary and the
radiation dominated eras of the cosmological expansion. With this
constrain it is

\begin{equation}
\alpha=1+i\frac{\sqrt{H_{ds}H_{0}}}{\omega}-\frac{H_{ds}H_{0}}{2\omega^{2}}\label{eq: alfa}\end{equation}

\begin{equation}
\beta=\frac{H_{ds}H_{0}}{2\omega^{2}}\label{eq: beta}\end{equation}

In eqs. (\ref{eq: alfa}) and (\ref{eq: beta}) $\omega=k/a(\eta_{0})$
is the angular frequency that would be observed today, and $H_{0}=1/\eta_{0}$
is the Hubble expansion rate that would be observed today. These calculations
are called \textit{Bogoliubov coefficient methods} \cite{key-7,key-12,key-13}. 

In inflationary scenarios both of classical and macroscopic perturbations
are damped out by inflation. Thus, the minimum allowed level of fluctuations
is that required by quantum uncertainty principle. The choice of the
solution (\ref{eq: ampiezza inflaz.}) corresponds precisely to such
a De Sitter vacuum state \cite{key-7,key-12,key-13}. Then, if the
period of inflation was long enough, the observable properties of
the Universe today should be indistinguishable from the properties
of a Universe started in the De Sitter vacuum state.

In the radiation dominated phase the eigenmodes which describe particles
are the coefficients of $\alpha$, and the eigenmodes which describe
antiparticles are the coefficients of $\beta$ \cite{key-7,key-12,key-13}.
Thus, the number of created particles of angular frequency $\omega$
in the radiation dominated phase is given by 

\begin{equation}
N_{\omega}=|\beta_{\omega}|^{2}=\left(\frac{H_{ds}H_{0}}{2\omega^{2}}\right)^{2}.\label{eq: numero quanti}\end{equation}

In this way, the expression for the energy density of the stochastic
relic gravitons background in the frequency interval $(\omega,\omega+d\omega)$
can be written down like

\begin{equation}
d\rho_{gw}=2\omega\left(\frac{\omega^{2}d\omega}{2\pi^{2}}\right)N_{\omega}=\frac{H_{ds}^{2}H_{0}^{2}}{4\pi^{2}}\frac{d\omega}{\omega}=\frac{H_{ds}^{2}H_{0}^{2}}{4\pi^{2}}\frac{df}{f}.\label{eq: de energia}\end{equation}

Eq. (\ref{eq: de energia}) can be re-written in terms of the present
day and the De Sitter energy-density of the universe. For the Hubble
expansion rates it is\begin{equation}
\begin{array}{c}
H_{0}^{2}=\frac{\rho_{c}}{3}\\
\\H_{ds}^{2}=\frac{\rho_{ds}}{3}.\end{array}\end{equation}
Then, the spectrum is given by \cite{key-7,key-12,key-13}

\begin{equation}
\Omega_{gw}(f)=\frac{1}{\rho_{c}}\frac{d\rho_{gw}}{d\ln f}=\frac{f}{\rho_{c}}\frac{d\rho_{gw}}{df}=\frac{16}{9}\frac{\rho_{ds}}{\rho_{Planck}},\label{eq: spettro gravitoni}\end{equation}

and the introduced Planck density $\rho_{Planck}$ is normalized in
our units.

Actually, the calculation works for a very simplified model that does
not include the matter dominated era. If this era is also included
the redshift has to be considered. An enlightening computation in
\cite{key-7} gives

\begin{equation}
\Omega_{gw}(f)=\frac{16}{9}\frac{\rho_{ds}}{\rho_{Planck}}(1+z_{eq})^{-1},\label{eq: spettro gravitoni redshiftato}\end{equation}
for the waves which at the time in which the Universe was becoming
matter dominated had a frequency higher than $H_{eq}$, the Hubble
constant at that time. This corresponds to frequencies $f>(1+z_{eq})^{1/2}H_{0}$,
where $z_{eq}$ is the redshift of the Universe when the matter and
radiation energy density were equal. The redshift correction in equation
(\ref{eq: spettro gravitoni redshiftato}) is needed because the Hubble
parameter, which is governed by Friedman equations, should be different
from the observed one $H_{0}$ for a Universe without matter dominated
era.

At lower frequencies the spectrum is given by \cite{key-7,key-12,key-13}

\begin{equation}
\Omega_{gw}(f)\thicksim f^{-2}.\label{eq: spettro basse frequenze}\end{equation}

The results (\ref{eq: spettro gravitoni}) and (\ref{eq: spettro gravitoni redshiftato})
cannot be applied in all the range of physical frequencies. In fact,
for waves with frequencies less than $H_{0}$ today, the notion of
energy density has no sense, because the wavelength becomes longer
than the scale of the Universe. In the same way, at high frequencies
there is a maximum frequency above which the spectrum drops to zero
rapidly \cite{key-7,key-12,key-13}. 

In the above calculation, the simple assumption that the phase transition
from the inflationary to the radiation dominated epoch is instantaneous
has been implicitly made. In the real Universe this process occurs
over some time scale $\Delta\tau$, and above a frequency

\begin{equation}
f_{max}=\frac{a(t_{1})}{a(t_{0})}\frac{1}{\Delta\tau},\label{eq: freq. max}\end{equation}

which is the red shifted rate of the transition, $\Omega_{gw}$ drops
rapidly. The two cutoffs (at low and high frequencies) to the spectrum
guarantee that the total energy density of the relic GWs is finite
\cite{key-7,key-12,key-13}.

\section{The conformal analysis by using the inflaton field}

Now, a variation of a recent treatment \cite{key-17}, that used a
conformal analysis, will be considered.

The GW-equations in the TT gauge are \cite{key-16}

\begin{equation}
\square h_{i}^{j}=0.\label{eq: dalembert}\end{equation}

Matter perturbations do not appear in (\ref{eq: dalembert}) since
scalar and tensor perturbations do not couple with tensor perturbations
in Einstein equations \cite{key-17}. 

The more general scalar-tensor action in 4 dimensions is given by
\cite{key-12} \begin{equation}
S=\int d^{4}x\sqrt{-g}[\varphi R-\frac{\omega(\varphi)}{\varphi}g^{\mu\nu}\varphi_{;\mu}\varphi_{;\nu}-W(\varphi)+\mathcal{L}_{m}],\label{eq: scalar-tensor2}\end{equation}
and in the framework of inflationary theories the scalar field $\varphi$
works like inflaton \cite{key-9,key-20}. 

One can perform the conformal transformation \cite{key-17} \begin{equation}
\tilde{g}_{\alpha\beta}=e^{2\Phi}g_{\alpha\beta}\label{eq: conforme}\end{equation}

where the conformal rescaling \begin{equation}
e^{2\Phi}=\varphi\label{eq: rescaling}\end{equation}

has been chosen. $\Phi$ is the {}``conformal scalar field''. The
difference with reference \cite{key-17} is that in such a work the
analysis concerned $f(R)$ theories, while we focus our attention
on the inflaton scalar field $\varphi$.

By applying the conformal transformation (\ref{eq: conforme}) to
the action (\ref{eq: scalar-tensor2}) the conformal equivalent Hilbert-Einstein
action \begin{equation}
A=\int {\frac{1}{2k}d^{4}x\sqrt{-\widetilde{g}}[\widetilde{R}+L(\Phi,\Phi_{;\alpha})]},\label{eq: conform}\end{equation}

is obtained. $L(\Phi,\Phi_{;\alpha})$ is the conformal scalar field
contribution derived from

\begin{equation}
\tilde{R}_{\alpha\beta}=R_{\alpha\beta}+2(\Phi_{;\alpha}\Phi_{;\beta}-g_{\alpha\beta}\Phi_{;\delta}\Phi^{;\delta}-\frac{1}{2}g_{\alpha\beta}\Phi^{;\delta}{}_{;\delta})\label{eq: conformRicci}\end{equation}

and \begin{equation}
\tilde{R}=e^{-2\Phi}(R-6\square\Phi-6\Phi_{;\delta}\Phi^{;\delta}).\label{eq: conformRicciScalar}\end{equation}
In the rescaled action (\ref{eq: conform}) the matter contributions
have note been considered because our interaction with GWs concern
the linearized theory in vacuum. 

In \cite{key-17} it has been shown that $h_{i}^{j}$ is a conformal
invariant and that the d'Alembert operator transforms as 

\begin{equation}
\widetilde{\square}=e^{-2\Phi}(\square+2\Phi^{;\alpha}\partial_{;\alpha}).\label{eq: quadratello}\end{equation}

Thus, the background is changing while the tensor wave amplitude is
fixed. 

In order to study the cosmological stochastic background, the operator
(\ref{eq: quadratello}) has to be specified for a Friedman-Robertson-Walker
metric \cite{key-17}, obtaining

\begin{equation}
\ddot{h}_{+}+(3H+2\dot{\Phi})\dot{h}_{+}+k^{2}a^{-2}h_{+}=0,\label{eq: evoluzione h}\end{equation}

being $\square=\frac{\partial}{\partial t^{2}}+3H\frac{\partial}{\partial t}$,
$a(t)$ the scale factor and $k$ the wave number. 

Considering the conformal time $d\eta=dt/a$, eq. (\ref{eq: evoluzione h})
reads\begin{equation}
\frac{d^{2}}{d\eta^{2}}h_{+}+2\frac{\gamma'}{\gamma}\frac{d}{d\eta}h_{+}+k^{2}h_{+}=0,\label{eq: evoluzione h 3}\end{equation}

where $\gamma=ae^{\Phi}$. Inflation means that $a(t)=a_{0}\exp(Ht)$
and then $\eta=\int dt/a=1/(aH)$ and $\frac{\gamma'}{\gamma}=-\frac{1}{\eta}.$
The exact solution of (\ref{eq: evoluzione h 3}) is \cite{key-17}
\begin{equation}
h_{+}(\eta)=k{}^{-3/2}\sqrt{2/k}[C_{1}(\sin k\eta-\cos k\eta)+C_{2}(\sin k\eta+\cos k\eta)].\label{eq: sol ev h2}\end{equation}

Inside the $1/H$ radius it is $k\eta\gg1.$ Furthermore, considering
the absence of GWs in the initial vacuum state, only negative-frequency
modes are present and then the adiabatic behavior is \cite{key-17}
\begin{equation}
h_{+}=k{}^{1/2}\sqrt{2/\pi}\frac{1}{aH}C\exp(-ik\eta).\label{eq: sol ev h3}\end{equation}

At the first horizon crossing ($aH=k$ at $t=10^{-22}$ second after
the Initial Singularity, see \cite{key-7}), the averaged amplitude
$A_{h}=(k/2\pi)^{3/2}h_{+}$ of the perturbations is \begin{equation}
A_{h}=\frac{1}{2\pi{}^{2}}C\label{eq: Ah}\end{equation}

when the scale $a/k$ grows larger than the Hubble radius $1/H,$
the growing mode of evolution is constant, i.e. {}``frozen'' \cite{key-17}.
This situation corresponds to the limit $-k\eta\ll1$ in equation
(\ref{eq: sol ev h2}). 

The amplitude $A_{h}$ of the wave is preserved until the second horizon
crossing after which it can be observed, in principle, as an anisotropy
perturbation of the CBR \cite{key-10,key-11}. It can be shown that
$\frac{\delta T}{T}\leq A_{h}$ is an upper limit to $A_{h}$ since
other effects can contribute to the background anisotropy \cite{key-18}.
Then, it is clear that the only relevant quantity is the initial amplitude
$C$ in equation (\ref{eq: sol ev h3}) which is conserved until the
re-enter. Such an amplitude directly depends on the fundamental mechanism
generating perturbations that depends on the inflaton scalar field
which generates inflation.

Considering a single monocromatic GW, its zero-point amplitude is
derived through the commutation relations \cite{key-17} \begin{equation}
[h_{+}(t,x),\pi_{h}(t,y)]=i\delta^{3}(x-y)\label{eq: commutare}\end{equation}

calculated at a fixed time $t.$ 

As it is \cite{key-17}

\begin{equation}
\pi_{h}=e^{2\Phi}a^{3}\dot{h}_{+},\label{eq: pi h}\end{equation}

equation (\ref{eq: commutare}) reads\begin{equation}
[h_{+}(t,x),\dot{h}_{+}(y,y)]=i\frac{\delta^{3}(x-y)}{e^{2\Phi}a^{3}}\label{eq: commutare 2}\end{equation}

and the fields $h_{+}$ and $\dot{h}_{+}$ can be expanded in terms
of creation and annihilation operators

\begin{equation}
h_{+}(t,x)=\frac{1}{(2\pi)^{3/2}}\int d^{3}k[h_{+}(t)e^{-ikx}+h_{+}^{*}(t)e^{ikx}]\label{eq: crea}\end{equation}
\begin{equation}
\dot{h}_{+}(t,x)=\frac{1}{(2\pi)^{3/2}}\int d^{3}k[\dot{h}_{+}(t)e^{-ikx}+\dot{h}_{+}^{*}(t)e^{ikx}].\label{eq: distruggi}\end{equation}

The commutation relations in conformal time are then \cite{key-17}
\begin{equation}
[h_{+}\frac{d}{d\eta}h_{+}^{*}-h_{+}^{*}\frac{d}{d\eta}h]=i\frac{8\pi^{3}}{e^{2\Phi}a^{3}}.\label{eq: commutare 3}\end{equation}

Inserting (\ref{eq: sol ev h3}) and (\ref{eq: Ah}), it is $C=\sqrt{2}\pi{}^{2}He^{-\Phi}$
where $H$ and $\Phi$ are calculated at the first horizon crossing
and then \begin{equation}
A_{h}=\frac{\sqrt{2}}{2}He^{-\Phi},\label{eq: Ah2}\end{equation}

which means that the amplitude of GWs produced during inflation directly
depends on the inflaton field being $\Phi=\frac{1}{2}\ln\varphi.$
Explicitly, it is \begin{equation}
A_{h}=\frac{H}{\sqrt{2\varphi}},\label{eq: Ah3}\end{equation}

thus, we have found a relation that links directly the amplitude of
relic GWs with the inflaton scalar field $\varphi$ which generates
inflation:

\begin{equation}
\varphi=\frac{H^{2}}{2A_{h}^{2}}.\label{eq: fi}\end{equation}

\section{The inflaton field and the slow roll condition on inflation}

It is well known that the requirement for inflation, which is $p=-\rho$
\cite{key-9,key-20}, can be approximately met if one requires $\dot{\varphi}<<V(\varphi)$,
where $(\varphi)$ is the potential density of the field. This leads
to the so called \textit{slow-roll approximation} (SRA), which provides
a natural condition for inflation to occur \cite{key-9,key-20}. The
constraint on $\dot{\varphi}$ is assured by requiring $\ddot{\varphi}$
to be negligible. With such a requirement, the slow-roll parameters
are defined (in natural units) by \cite{key-9,key-20}

\begin{equation}
\begin{array}{c}
\epsilon(\varphi)\equiv\frac{1}{2}(\frac{V'(\varphi)}{V(\varphi)})^{2}\\
\\\eta(\varphi)\equiv\frac{V''(\varphi)}{V(\varphi)}.\end{array}\label{eq: slow-roll}\end{equation}

Then, the SRA requirements are \cite{key-9,key-20}:

\begin{equation}
\begin{array}{c}
\epsilon\ll1\\
\\|\eta|\ll1,\end{array}\label{eq: slow-roll2}\end{equation}

that are satisfied when it is \cite{key-9,key-20}

\begin{equation}
\varphi\gg M_{Planck},\label{eq: planckiano}\end{equation}

where the Planck mass, which is $M_{Planck}\simeq2.177*10^{-5}g$
in ordinary units and $M_{Planck}=1$ in natural units \cite{key-16},
has been introduced \cite{key-9,key-20}.

Now, by using a connection between the two treatments on relic GWs,
i.e. the traditional one and the conformal one, we show that the condition
(\ref{eq: planckiano}) is satisfied.

Let us start by recalling the equation for the characteristic amplitude
$h_{C}$, see Equation 65 in \cite{key-19}, 

\begin{equation}
h_{C}(f)\simeq1.26*10^{-18}\left(\frac{1{\rm Hz}}{f}\right)\sqrt{h_{100}^{2}\Omega_{gw}(f)}.\label{eq: legame ampiezza-spettro}\end{equation}

This equation gives a value of the amplitude of the relic GWs stochastic
background in function of the spectrum in the frequency range of ground
based detectors \cite{key-19}. Such a amplitude is also the strain
applied on the detector's arms \cite{key-19}. Such a range is given
by the interval $10Hz\leq f\leq10KHz$ \cite{key-1}.

Using eq. (\ref{eq: spettro gravitoni}) eq. (\ref{eq: legame ampiezza-spettro})
becomes

\begin{equation}
h_{C}(f)\simeq1.26*10^{-18}\left(\frac{1{\rm Hz}}{f}\right)\sqrt{h_{100}^{2}\frac{16}{9}\frac{\rho_{ds}}{\rho_{Planck}}}.\label{eq: hc}\end{equation}

The mean value of this quantity will be $\simeq A_{h},$ thus, by
using eq. (\ref{eq: Ah3}) it is

\begin{equation}
\frac{H}{\sqrt{2\varphi}}\simeq\frac{1.26*10^{-18}\sqrt{h_{100}^{2}\frac{16}{9}\frac{\rho_{ds}}{\rho_{Planck}}}\int_{10}^{10000}f^{-1}df}{\int_{10}^{10000}df},\label{eq: integra}\end{equation}

that, by computing the integrals in the range $10Hz\leq f\leq10KHz$
and recalling that $h_{100}\thickapprox0.74$ \cite{key-9,key-10,key-11}
and that for GUT energy-scale of inflation it is \cite{key-7,key-12,key-13}

\begin{equation}
\frac{\rho_{ds}}{\rho_{Planck}}\approx10^{-12},\label{eq: rapporto densita primordiali}\end{equation}
gives \begin{equation}
\frac{H}{\sqrt{2\varphi}}\simeq8.2*10^{-28}.\label{eq: integrato}\end{equation}

By restoring ordinary units and recalling that $H\simeq10^{22}Hz$
at the first horizon crossing \cite{key-7}, at the end it is \begin{equation}
\varphi\simeq7*10^{4}g.\label{eq: inflaton value}\end{equation}

This value agrees with the slow roll condition on inflation. In fact,
the condition (\ref{eq: planckiano}) is surely satisfied being $M_{Planck}\simeq2.177*10^{-5}g$
in ordinary units.

\section{Conclusions}

After reviewing a traditional analysis on the relic GWs spectrum,
in this paper a variation a of a more recent analysis that uses a
conformal treatment has been discussed. A connection between the two
different treatments has been analysed too. This connection permitted
to obtain an interesting equation for the inflaton field. This equation
gived a value that agrees with the slow roll condition on inflation.

\section*{Acknowledgements}

The Associazione Scientifica Galileo Galilei has to be thanked for
supporting this paper. I strongly thank the referee for the good interpretation
of the results and for correcting some misprints.

\end{document}